\documentstyle[pre,aps,epsf]{revtex}

\begin{document}

\draft

%
%
%
\title
{Exact multipoint and multitime correlation functions of a one-dimensional
 model of adsorption and evaporation of dimers}

\author
{Mauro Mobilia }

\address
{Institute of Theoretical Physics, Swiss Federal Institute of Technology of  Lausanne, CH-1015 Lausanne  EPFL, Switzerland }

\date{\today}

\maketitle

\begin{abstract}
 In this work, we provide a method which allows to compute exactly the 
multipoint and multi-time correlation functions of a one-dimensional
stochastic model of dimer adsorption-evaporation with random (uncorrelated) initial states.
 In particular explicit expressions of the two-point noninstantaneous/instantaneous 
correlation functions are obtained.
The long-time behavior of these expressions is discussed in details and
in various physical regimes.
\end{abstract}
One-dimensional reaction-diffusion (RD) processes of {\it interacting particles} have been
 extensively studied in the last decade
because of their relevance as  examples of strongly-correlated nonequilibrium systems and their 
connexion with experimental situations \cite{Privman,Kuroda,Vilfan,Schutzrev}. 

Among the RD systems, the ``diffusion-limited with pair-annihilation and creation'' (DPAC)
 model
 \cite{Lushnikov1,Grynberg1-corr,Schutz2,Schutz1,Santos1,Mobar2,Grynberg-auto,Barmo1,Oliveira,Barmo2,Torney} 
plays a particular role. In fact it is one of the rare nonequilibrium model for which it has been possible, in some special cases, to compute some
 {\it dynamical correlation functions}. In addition this model carries valuable informations
 for various experimental situations where particles diffuse and dimer can be adsorpted/evaporated 
\cite{Privman,Kuroda}.
Despite the interest in the DPAC-model, not all the desirable information
 on the correlation functions
was available so far. In particular considerably less results are available
for the (complete) DPAC-model than for  the ``diffusion-limited pair annihilation'' (DPA) model
(where there is no pair-creation).

 Recently, there has been a regain of interest for the  study of the  DPAC-model because
 its possible application in  various fields
such as the experimental study
 of the photogrowth properties of long-lived midgap absorption band in a $MX$ chain \cite{Kuroda}  and
 in interdisciplinary studies \cite{Vilfan}. In particular, it  has been shown that the autocorrelation functions of the DPAC-model provide valuable informations on the relaxation of biological dimer adsorption \cite{Vilfan}. In this work we consider the (free-fermion) DPAC and DPA-model and obtain results which remained
 inaccessible so far. In particular, we explicitly compute the exact and complete expression of the noninstantaneous two-point correlation functions for random initial conditions and then analyse the 
long-time behavior of the latter.

We consider a periodic lattice of $L$ sites (without restriction, $L$ is assumed to be {\it even})
on which an even number of (classical) particles interact. Each site is either empty or occupied by a particle at most
(because of the {\it hard-core interaction}). When a particle and a vacancy are adjacent
 to each other, the particle can {\it jump} to the right with a rate $h'$ or to the left with
 a rate $h$. When two particles are adjacent, they can {\it annihilate in pairs} with a rate $\epsilon$. In addition, when two vacancy are adjacent, a {\it pair of particles can be created} with rate $\epsilon'$.
We now adopt the so-called {\it stochastic Hamiltonian} formalism (see e.g. \cite{Schutzrev} and references
 therein). To do this,  at each of the 
$L-$lattice sites, we associate to a particle (vacancy) a spin-${\frac{1}{2}}$ down (up) . In so doing the master equation governing the dynamics of the model
can formally be rewritten as an imaginary-time Schr\"odinger equation for a quantum spin-chain:
 $ (\partial / \partial t) |P(t)\rangle = - H|P(t)\rangle$, where $|P(t)\rangle =\sum_{\{n\}}
P(\{n\},t)|\{n\}\rangle$ describes the state of the system at time $t$ 
(the sum runs over all the $2^L$
configurations $\{n\}$). Performing a standard Jordan-Wigner transformation  
and then a Fourier transformation, the {\it stochastic Hamiltonian} can be recasted in a 
{\it fermionic representation}  \cite{Schutzrev,Lushnikov1,Grynberg1-corr,Schutz2,Schutz1,Santos1,Mobar2,Grynberg-auto,Barmo1}. We also define the ``left vacuum'' $\langle {\widetilde \chi}| \equiv \sum_{\{n\}}
\langle \{ n\}|$. The probability conservation yields $\langle {\widetilde \chi}|H=0$.

Exact solution of the DPAC-model is only possible in the {\it free-fermion case}
\cite{Schutzrev,Lushnikov1,Grynberg1-corr,Schutz2,Schutz1,Santos1,Mobar2,Grynberg-auto,Barmo1}
, and therefore, with $\gamma \equiv \epsilon +\epsilon'-(h+h')$,  one has to impose: $\gamma=0$. 

Hereafter, we always consider that the constraint $\gamma =0$ is fulfilled. In this case, the stochastic 
Hamiltonian reads:
\begin{eqnarray}
\label{eq.5.1}
H&=&\sum_{q>0} \left[\omega(q) a_q^{\dagger}a_q +\ 
\omega^{\ast}(q) a_{-q}^{\dagger}a_{-q} +2\sin{q}
\left(\epsilon a_q a_{-q} + \epsilon' a_{-q}^{\dagger}
a_{q}^{\dagger}\right)\right]+\epsilon' L
\end{eqnarray}  
where $a_q^{\dagger}$ and $a_q$ are usual fermion operators. In addition, $\omega(q)\equiv{\widetilde c}-b\cos{q}+iv\sin{q}$, with $b\equiv \epsilon+\epsilon'$,
 $\widetilde{c}\equiv \epsilon- \epsilon'$, 
$v\equiv h'-h$ and $q=\pm \pi(2l-1)/L, \;\; l=1, 2, \dots, L/2$.

We consider 
translationally invariant and uncorrelated random initial states $|\rho_{0}\rangle$ 
with an even number $N$ of particles, of density $\rho_{0}=N/L$, 
i.e. $|\rho_0\rangle=
\left(
 \begin{array}{c}
 1-\rho_{0}    \\
  \rho_{0}
\end{array}\right)^{\otimes L}$. From the fermion reformulation, an important property of the left vacuum follows (in the subspace with an even number
 of particles that we consider here):
 $\langle \widetilde \chi| a_q^{\dagger}=\cot(\frac{q}{2})\langle \widetilde \chi| a_{-q}$, 
\cite{Schutzrev,Schutz1,Santos1}.

The  {\it free-fermion} character of the stochastic Hamiltonian (\ref{eq.5.1}) allows the computation of the following zero-time  correlators \cite{Santos1}, with $\mu\equiv \rho_0/(1-\rho_0)$:
\begin{eqnarray}
\label{eq.5.7.2}
\langle \widetilde{\chi}|a_{q'}a_q  |\rho_0\rangle &\equiv&\langle a_{q'}a_q \rangle(0)= 
\frac{ \mu^2 \cot{(q/2)} }{1+\mu^2 \cot^{2}{(q/2)}} \delta_{q,-q'}  \nonumber\\
\langle a_{q_1} a_{q_2}\dots a_{q_{2n-1}} a_{q_{2n}}\rangle(0)&=&
\frac{1}{n!}\sum_{\pi} {\cal S}(\pi) \langle a_{q_{\pi(1)} } a_{q_{\pi(2)} } \rangle(0)
\dots \langle a_{q_{\pi(2n-1)} } a_{q_{\pi(2n)} } \rangle(0),
\end{eqnarray}
where the sum is over all the permutations $\pi$ of the indices $\{q_1, q_2, \dots, q_{2n}\}$, 
with the constraints $\pi(1)<\pi(2); \dots; \pi(2n-1)<\pi(2n)$. Each
 permutation $\pi$ has a 
signature ${\cal S}(\pi)$ . 

It is advantageous at this point to introduce the ``pseudo-fermion'' operators \cite{Grynberg1-corr}
$\xi_{q} = \alpha^{-1} \cos{\theta_q} a_q + \alpha \sin{\theta_q} a_{-q}^{\dagger}  $ and $
\xi_{q}^+ = \alpha \cos{\theta_q} a_q^{\dagger} + \alpha^{-1} \sin{\theta_q} 
a_{-q}\;$, with 
 $\theta_{-q}=-\theta_q $. Although they are  {\it not adjoint} each other \cite{Grynberg1-corr}, these
 operators fulfill the canonical anticommutation relation $\{\xi_{q},\xi_{q'}^+ \}=
\delta_{q,q'}$. The probability conservation yields $\langle \widetilde{\chi}|\xi_{q}^{+} =0$ \cite{Grynberg1-corr}.

It has been shown \cite{Grynberg1-corr} that if one chooses
the following parameters
 $\tan (2\theta_{q})=\frac{2\sqrt{\epsilon \epsilon'}\sin{q}}{b\cos{q}-\widetilde{c}} \:\; ;\;\;
\alpha^2 = \sqrt{\epsilon'/\epsilon},$
the stochastic free-fermion Hamiltonian (\ref{eq.5.1}) is {\it diagonal} in the
pseudo-fermion representation and reads: $H=\sum_q \lambda_q \xi_q^+ \xi_q,
 \;\; \mbox{ with}\;\;  \lambda_q = b-{\widetilde c}\cos{q} +i v\sin{q} $. 
Because of this diagonal representation, the pseudo-fermion operators 
 evolve according to $\xi_{q}(t) = {\rm e}^{-\lambda_q t}\xi_{q}(0) $ and $\xi_{q}^+(t) =
 {\rm e}^{\lambda_q t}\xi_{q}^+(0) $.

Using the fact that,  the pseudo-fermion operators $\xi_{q}$ are
{\it linear combination} of fermion operators $a_q$ and $a_q^{\dagger}$, and using the expression
(\ref{eq.5.7.2})  as well as the property of the left vacuum, it follows that:
\begin{eqnarray}
\label{eq.5.7.7}
 \langle \xi_{q} \xi_{q'} \rangle(0) =
\frac{\cos\theta_q \cos\theta_{q'}}{\alpha^2} \langle a_{q} a_{q'} \rangle(0)
+ \alpha^{2}\sin\theta_q \sin\theta_{q'} \langle a_{-q}^{\dagger} a_{-q'}^{\dagger} \rangle(0)
+ \sin\theta_{q'}\cos\theta_q \langle a_{q} a_{-q'}^{\dagger} \rangle(0)
+\sin\theta_{q}\cos\theta_{q'} \langle a_{-q}^{\dagger} a_{q'} \rangle(0)
\end{eqnarray}
For the sequel it is useful to compute $\langle \xi_q \xi_{q'} 
\rangle(0)\equiv
\langle {\widetilde \chi}|\xi_q \xi_{-q}|\rho_0\rangle \delta_{q', -q} 
$. Therefore we introduce the following function :
\begin{eqnarray}
\label{eq.5.7.11.0}
 f(q) &\equiv& \langle \xi_q \xi_{-q}\rangle(0)=
\nu_q -\frac{\mu^{2}\epsilon \nu_q \cos^{2}(q/2)}{\left(1+\frac{\mu^2 \epsilon}{\epsilon'}\nu_q^2\right){\rm Re}(\lambda_q)} \left[
1+2\nu_q^{2}+2/\nu_q^{2} \right], \mbox{ where}\; \nu_q \equiv\sqrt{\epsilon'/\epsilon}\cot{(q/2)},
\end{eqnarray}

One can also check that
\begin{eqnarray}
\label{eq.5.7.9}
\langle  \xi_{q_{1}} \dots \xi_{q_{2n}}\rangle(0)
=\frac{1}{n!}\sum_{\pi} {\cal S}(\pi) \langle \xi_{q_{\pi(1)} }
 \xi_{q_{\pi(2)} } \rangle(0)
\dots \langle \xi_{q_{\pi(2n-1)} } \xi_{q_{\pi(2n)} } \rangle(0),
\end{eqnarray}
where we adopted the same notations as in (\ref{eq.5.7.2}).

\vspace{0.2cm}

Let us now sketch a four-step procedure which allows to
 compute explicitly the {\it multipoint} and {\it multi-time}
 correlation functions $\langle n_{j_1}(t_1)  \dots  n_{j_{m-1}}(t_{m-1}) n_{j_m}(t_m)
 \rangle \equiv 
\langle {\widetilde \chi}| n_{j_1}{\rm e}^{-H(t_1-t_2)} \dots n_{j_{m-1}}{\rm e}^{-H(t_{m-1}-t_{m})}
n_{j_m} {\rm e}^{-Ht_m}|\rho_0\rangle $:

\vspace{0.3cm}

(i) One first has to write the expression of the correlation functions in the
 Fourier space:
\begin{eqnarray}
\label{eq.5.7.10}
&& \langle n_{j_{1}}(t_1)  \dots  n_{j_{m}}(t_m) \rangle 
=\sum_{q_1,q_{1}',\dots,q_m,q_{m}'}{\rm e}^{i(q_{1}-q_{1}' )j_1 +
 \dots + i( q_{m}-q_{m}')j_m}
\langle a_{q_1}^{\dagger}(t_1) a_{q_{1}'}(t_1) 
\dots a_{q_m}^{\dagger}(t_m) a_{q_{m}'}(t_m)  \rangle.
\end{eqnarray}

(ii) One has then to rewrite the expression (\ref{eq.5.7.10}) in the 
{\it pseudo-fermion} language. This achieved with help of
\begin{eqnarray}
\label{eq.5.7.10.0}
a_{q}^{\dagger} a_{q'}=
\cos{\theta_q}\cos{\theta_{q'}} \xi_{q}^{+}\xi_{q'}
- \cos{\theta_q}\sin{\theta_{q'}} \xi_{q}^{+}\xi_{-q'}^{+}
-  \sin{\theta_q} \cos{\theta_{q'}} \xi_{-q}\xi_{-q'}
+  \sin{\theta_q}\sin{\theta_{q'}} \xi_{-q}\xi_{-q'}^{+}.
\end{eqnarray}

(iii) Using the fact that $H$ is {\it diagonal} in the 
(pseudo-fermion) representation [with (\ref{eq.5.7.10.0})], one extracts
 the time-dependence of terms appearing in the pseudo-fermion  rewritten expression (\ref{eq.5.7.10}). As
 an example, we have:
\begin{eqnarray}
\label{eq.5.7.10.1}
 \langle \xi_{q_1}(t_1) \xi_{q_1'}(t_1) \dots \xi_{q_m}(t_m)
 \xi_{q_m'}(t_m) \rangle  =
{\rm e}^{ -(\lambda_{q_1}+\lambda_{q_1'})t_1 -\dots 
-(\lambda_{q_m}+\lambda_{q_m'})t_m }
\langle \xi_{q_1} \xi_{q_1'} \dots \xi_{q_m}
 \xi_{q_m'} \rangle (0)
\end{eqnarray}  

(iv) Finally, the zero-time correlation functions of pseudo-fermion operators appearing in the
r.h.s. of (\ref{eq.5.7.10}), after the steps (i)-(iii), are computed 
with help of the Wick factorization (\ref{eq.5.7.9}) and using $\langle \widetilde{\chi}|\xi_{q}^{+}=0$.

\vspace{0.3cm}

This general four-step procedure 
provides a systematic method to obtain explicitly, starting from homogeneous
 random initial conditions,
the  multipoint and multitime 
correlation functions of the free-fermion model under consideration here.
It is however important to notice that the computation of each of these quantities 
leads to rather complicated technical difficulties.
Let us mention that using the domain-wall duality and with help of the generating function studied  in \cite{Mobar2}, we can compute the  stationary multipoint correlation functions
of the DPAC-model from the spin-spin correlation functions of the one-dimensional Ising model with a generalized (biased) Glauber's dynamics \cite{Grynberg1-corr,Mobar2,Mobar3}.
 In fact we can show  ($j_m> \dots >j_1$) that $\langle n_{j_1}
\dots n_{j_m}\rangle(\infty)=[\rho(\infty)]^{m}$ \cite{Mobar3}, where $\rho(\infty)=\sqrt{\epsilon'}/(\sqrt{\epsilon}+\sqrt{\epsilon'})$
is the stationary density of particles. In addition, for an homogeneous system with initial density
$\rho_0=1/2$ of particles, because of the quadratic form of the generating function,
 the expressions of  
spin-spin correlation functions of the dual of the DPAC-model are Pfaffians \cite{Mobar2,Mobar3}. In this case it is therefore possible, via the domain-wall duality 
\cite{Schutzrev,Mobar2}, to sort out the technical complications and
 explicitly
 compute the instantaneous multipoint correlation functions. As an example for the DPA-model ($\epsilon=h+h'$, $\epsilon'=0$), the long-time behavior ($bt\gg 1$) of the  three-point correlation functions read $8\langle n_{j}(t) n_{j+r_1}(t)n_{j+r_2}(t)\rangle\approx
\frac{1+8\{r_1^2 +r_1r_2(5(r_2-r_1)-1) \}}{1280\pi({\widetilde c}t)^{3}}$, with $r_2>r_1$ \cite{Mobar3}.    

Because of the general technical problems inherent to the computation  of the correlation functions, the above-mentioned systematic
four-step procedure is in particular useful to take into account random initial conditions, which affect the long-time
 dynamics of
the non-universal relaxation (when all the rates $h, h',\epsilon$ and $\epsilon'$ are $>0$, see below).

To illustrate  the difficulties which appear in computing the
multipoint and multitime correlation functions (from uncorrelated but random initial states), as well as their importance,
 one can point out the work of Derrida
and Zeitak \cite{Derrida1},  where 
these authors obtained the universal distribution of domain sizes of 
one-dimensional Potts model with zero temperature Glauber dynamics. 
This was achieved using the properties of coalescing random 
walkers to compute the probability of having the same value at time $t$
 for $N$ spins located at $N$ distinct and ordered sites, which is related 
to the distribution of domain sizes.
The authors also studied the domain-walls dynamics and thus
considered the following (free-fermion) RD model $A\emptyset\leftrightarrow\emptyset A,\;$
 $AA\rightarrow A$ and $AA\rightarrow \emptyset$, with reaction-rates $1$,
${\rm(q-2)/ (q-1)}$ and ${\rm 1/(q-1)}$, respectively. For this RD model, with random 
(but uncorrelated) initial conditions, the authors of 
\cite{Derrida1} computed
the density of particles and the instantaneous two-point correlation functions. 
It has to be noticed that there exists a similarity transformation (see e.g.
 \cite{Schutzrev} and references
 therein) which maps the DPA-model
studied  here (with $\epsilon'=0$) onto the RD model considered in 
\cite{Derrida1}.

Another important problem where the (stationary) multipoint correlation functions of a
free-fermion model play a relevant  role is the computation, for the q-state Potts model
 in the zero-temperature Glauber dynamics, of the exact
 persistence exponent which gives the fraction of spins $\widetilde{\rho}_L({\rm q})$
 which have never
flipped \cite{Derrida2,Derrida3}.
To compute $ \widetilde{\rho}_L({\rm q}) $, the authors mapped the problem onto an exactly 
solvable (free-fermion) RD model:  $A \emptyset\leftrightarrow \emptyset A$ (with reaction-rate $1$) and
$AA\rightarrow A$ (with reaction-rate $2$). In addition it is assumed that a ``source'' ensures that the
origin of the (periodic) lattice is always occupied.
Starting from an uncorrelated but random initial state 
(with the initial site always occupied) denoted $|P''(0)\rangle$ , the 
problem of finding   $\widetilde{\rho}_L({\rm q})$ reduces to the
computation of the following multipoint correlator
\cite{Derrida2} 
:
$\widetilde{\rho}_L({\rm q})= {\rm lim_{t\rightarrow \infty}} \langle 0|(1+a_L) \dots
(1+a_1)e^{-H't}| P''(0)\rangle $, where the $a_j$'s are fermion operators, $|0\rangle$
is the vacuum ($a_j| 0\rangle=0$) and $H'$ is the stochastic Hamiltonian associated
to RD model considered in \cite{Derrida2,Derrida3}.

\vspace{0.5cm}

To be specific we now focus on the computation
of the connected non-instantaneous two-point correlation functions of 
the DPAC-model for random initial conditions $|\rho_0\rangle$. 

Following the above-mentioned four-step procedure [(i)-(iv)], 
adopting the notation $\phi_q\equiv {\rm Re}(\lambda_q)=b-{\widetilde c}\cos{q}$, 
we obtain, in the thermodynamic limit ($L\rightarrow \infty$):
\begin{eqnarray}
\label{eq.5.7.11}
{\cal C}_r(t,t_0)&\equiv&   \langle n_{j+r}(t+t_0) n_j (t_0)\rangle -\rho(t+t_0)\rho(t_0)
 =
\epsilon \epsilon'\left( \int_{0}^{\pi} \frac{dq}{\pi}
 \sin[qr-vt\sin{q}]\frac{\sin{q}}{\phi_q} {\rm e}^{-\phi_q t}\right)^{2}
\nonumber\\
&+& 4\epsilon\epsilon' \left( \int_{0}^{\pi} \frac{dq}{\pi}
 \cos[qr-vt\sin{q}]\frac{\sin^{2}(q/2)}{\phi_q} {\rm e}^{-\phi_q t}\right)\left(
 \int_{0}^{\pi} \frac{dq}{\pi}
 \cos[qr-vt\sin{q}]\frac{\cos^{2}(q/2) }{\phi_q} {\rm e}^{-\phi_q t}
\right)\nonumber\\
&+& 2\epsilon'\sqrt{\epsilon\epsilon'} \left( \int_{0}^{\pi} \frac{dq}{\pi}
 \sin[qr-vt\sin{q}]\frac{\sin{q}}{\phi_q} {\rm e}^{-\phi_q t}\right)\left(
 \int_{0}^{\pi} \frac{dq}{\pi}
 \sin[qr-vt\sin{q}]\frac{\cos^2{(q/2)}}{\phi_q} f(q){\rm e}^{-\phi_q (t+2t_0)}
\right)\nonumber\\
&-& \sqrt{\epsilon\epsilon'} \left( \int_{0}^{\pi} \frac{dq}{\pi}
 \cos[qr-vt\sin{q}]\frac{{\widetilde c}-b\cos{q}}{\phi_q} 
{\rm e}^{-\phi_q t}\right)\left(
 \int_{0}^{\pi} \frac{dq}{\pi}
 \cos[qr-vt\sin{q}]\frac{\sin{q}}{\phi_q} f(q){\rm e}^{-\phi_q (t+2t_0)}
\right)\nonumber\\
&-& 2\epsilon\sqrt{\epsilon\epsilon'} \left( \int_{0}^{\pi}
 \frac{dq}{\pi}
 \sin[qr-vt\sin{q}]\frac{\sin{q}}{\phi_q}
 {\rm e}^{-\phi_q t}\right)\left(
 \int_{0}^{\pi} \frac{dq}{\pi}
 \sin[qr-vt\sin{q}]\frac{\sin^{2}{(q/2)}}{\phi_q} f(q){\rm e}^{-\phi_q (t+2t_0)}
\right)\nonumber\\
&-&4\epsilon\epsilon'  \left( \int_{0}^{\pi} \frac{dq}{\pi}
 \sin[qr-vt\sin{q}]\frac{\cos^{2}{(q/2)}}{\phi_q} f(q) {\rm e}^{-\phi_q (t+2t_0)}\right) 
 \left( \int_{0}^{\pi} \frac{dq}{\pi}
 \sin[qr-vt\sin{q}]\frac{\sin^{2}{(q/2)}}{\phi_q} f(q) {\rm e}^{-\phi_q (t+2t_0)}\right)\nonumber\\
&-& \epsilon\epsilon'
 \left( \int_{0}^{\pi} \frac{dq}{\pi}
 \cos[qr-vt\sin{q}]\frac{\sin{q}}{\phi_q} f(q) {\rm e}^{-\phi_q (t+2t_0)}\right)^{2}
\end{eqnarray}

In this expression $\rho(t)=\frac{\sqrt{\epsilon'}}{\sqrt{\epsilon} + 
\sqrt{\epsilon'} }  - \sqrt{\epsilon \epsilon'}\int_{0}^{\pi} \frac{dq}{\pi} 
\frac{\sin{q}}{\phi_q}
 f(q){\rm e}^{-2\phi_q t} $  designates the (translationally-invariant) density of particles at time $t$. The latter has been previously 
studied (for the initial states $|\rho_0\rangle$) in \cite{Mobar2}.

The equation (\ref{eq.5.7.11}) is the main result of this work and provides the complete 
expression of the  noninstantaneous two-point correlation functions of the (free-fermion version of the)
 DPAC-model. From the latter, it is clear that
 one can also obtain the {\it instantaneous} two-point
 connected  correlation functions ${\cal C}_{r}(t=0,t_0>0)$, and  one can 
check , as expected from the general properties of the DPAC-model \cite{Schutz1},
 that the latter do not depend on the bias $v$.
In addition, it is clear that the expression (\ref{eq.5.7.11}) also includes  the instantaneous and  noninstantaneous  connected two-point
correlation functions of the DPA-model, where $\epsilon'=0$ and $b={\widetilde c}>0$.
 It is worthwhile to notice that the noninstantaneous correlation functions
(\ref{eq.5.7.11})  [with $t>0$ and $v\neq 0$] depend on the {\it sign} of $r$
 :  ${\cal C}_{r}(t,t_0; v)={\cal C}_{-r}(t,t_0; -v)$.
 Conversely, the {\it instantaneous} correlation functions [with $t=0$ and $t_0>0$ in
 (\ref{eq.5.7.11}) ] do  {\it not } depend on the {\it sign} of $r$.
Let us also stress the fact that setting in (\ref{eq.5.7.11}) $\mu=\infty$ (i.e. $\rho_0=1$), $\mu=0$
(i.e. $\rho_0=0$) or $\mu=1$ (i.e. $\rho_0=1/2$), we recover  results obtained 
in  \cite{Grynberg1-corr,Schutz2,Schutz1,Santos1,Mobar2,Grynberg-auto,Barmo1}.

To proceed with a long-time study of the expression
 (\ref{eq.5.7.11}), we have carried out a systematic
 asymptotic expansion of the integrals appearing in (\ref{eq.5.7.11}), in which the
small $q$ regime of integration dominates.
Hereafter, we analyse two different regimes
and distinguish the case with pair creation (i.e. with $\epsilon'>0$ and $\widetilde{c}\neq 0$) from the case with
only (asymmetric) diffusion and pair-annihilation (i.e. with $\epsilon'=0$
 and $\epsilon>0$). The cases where $\widetilde{c}=0 $ (i.e. $\epsilon=
\epsilon'$,  $h=h'$; and $\epsilon=\epsilon'$,  $h+h'=2\epsilon$)
 have already been studied
in \cite{Grynberg1-corr}.

\vspace{0.3cm}

1.i) We first consider the case where $\epsilon'>0$ and $\epsilon>0$ in the
regime where $\epsilon t, \epsilon t_0 \gg 1$ and $ \epsilon 't, \epsilon 't_0 \gg 1$ 
[with ${\widetilde c}>0$].
In this situation the main contribution to the noninstantaneous correlation function
arises from the second  term of the r.h.s. of (\ref{eq.5.7.11}).
 We thus obtain ($v\neq 0$, ${\widetilde c}>0$):
\begin{eqnarray}
\label{eq.5.7.17}
{\cal C}_{r}(t,t_0)\approx \frac{\epsilon(1-u){\rm e}^{-u-4\epsilon't}}{16\pi \epsilon' (\widetilde{c}t)^2}, \;\; u\equiv (r-vt)^2/{\widetilde c}t
\end{eqnarray}
It is clear from (\ref{eq.5.7.17}) that in this regime the late behavior 
of the noninstantaneous correlation functions 
${\cal C}_{r}(t,t_0)$ only depends on the time $t$ (and not on $t_0$). We
 notice the non-trivial effect of the bias $ v\neq 0$ through the parameter $u$.
In the absence of the bias  and for  $r<\infty$ (i.e.  for the
 autocorrelation functions) we obtain:
${\cal C}_{r}(t,t_0;v=0)\approx \frac{\epsilon}{32 \pi \epsilon'} 
\frac{{\rm exp}[-4\epsilon't ]}{({\widetilde c}t)^2}$.

In this regime we now focus of the long-time behavior of the
{\it instantaneous} correlation functions ${\cal C}_{r}(t)$ 
(obtained setting $t=0$ in (\ref{eq.5.7.11})
and relabelling $t_0$ as the  variable $t$). The cases
 $\rho_0=1$ and $\rho_0=0$ having
 been studied previously [${\cal C}_{r}(t; \rho_0=0,1)\propto 
{\rm e}^{-4\epsilon' t} t^{-\nu'}$, $\nu'=3/2$ for $\rho_0 =1$ and $\nu'=1/2$ 
for $\rho_0 =0$ \cite{Grynberg1-corr,Barmo1}], here we
 focus on the case of random initial states, i.e.  with $0<\mu<\infty$
 [and $\rho_0\neq \rho(\infty)=\sqrt{\epsilon'}/
(\sqrt{\epsilon'}+\sqrt{\epsilon})$, which would correspond to the {\it trivial
 case} where $f(q)\equiv 0$].
The main contribution to ${\cal C}_{r}(t)$ comes from the
third and fourth term of the r.h.s. of (\ref{eq.5.7.11}).
 Introducing the parameters:
\begin{eqnarray}
\label{eq.5.7.18}
4A_0 &\equiv& -\sqrt{\epsilon/\epsilon'}(1-\zeta)^2 \zeta^{r-1} \; ; \;
B_0 \equiv  -(1-\zeta^2)  \zeta^{r-1}\; ; \;
4C_0 \equiv \sqrt{\epsilon'/\epsilon}(1+\zeta)^2 \zeta^{r-1} \; ; \;
\zeta\equiv\frac{\sqrt{\epsilon}-\sqrt{\epsilon'}}
{\sqrt{\epsilon} +\sqrt{\epsilon'}} ,
\end{eqnarray}
we obtain ($r<\infty$):
\begin{eqnarray}
\label{eq.5.7.19}
{\cal C}_{r}(t)\approx 
\frac{\pi {\rm e}^{-4\epsilon't}
 {\rm {\cal F}(\mu, r,\epsilon,\epsilon')}}{4(\pi {\widetilde c}t)^{3/2}}
,
\end{eqnarray}
where the rather complicated expression of the amplitude ${\cal F}(\mu,r,\epsilon,\epsilon')$ 
reads:
\begin{eqnarray}
\label{eq.5.7.20}
{\cal F}(\mu,r,\epsilon,\epsilon')&=&\frac{A_0 -C_0}{4\mu^{2}\sqrt{\epsilon\epsilon'^5}}\left\{
2\mu^{2}(\epsilon \epsilon')^{3/2} -2 \epsilon'^{2}\sqrt{\epsilon \epsilon'} (\mu r)^{2}
+\epsilon\epsilon' \left( \sqrt{\epsilon'^3/\epsilon} + \mu^{2}[2 r^{2}
\sqrt{\epsilon'^3/\epsilon} -
3\sqrt{\epsilon \epsilon'}   ]\right)
\right\} \nonumber\\
&-&
\frac{B_0 r}{12\epsilon \epsilon'^{2}\mu^{2}}
\left\{
6\mu^{2}(\epsilon \epsilon')^{3/2} -\mu^{2}(2r^{2}+1)
\epsilon'^{2}\sqrt{\epsilon \epsilon'}
+ \epsilon \epsilon' \left( 3\sqrt{\epsilon'^3/\epsilon} +\mu^2 \left[
(2r^2 +1)\sqrt{\epsilon'^3 / \epsilon} -9\sqrt{\epsilon \epsilon'}
\right] \right)
\right\}
\end{eqnarray}

It is remarkable that conversely to the {\it instantaneous} correlation functions 
(\ref{eq.5.7.19}), which amplitude (\ref{eq.5.7.20}) depends on the initial state through
 the parameter $0<\mu <\infty$, the long-time behavior of the noninstantaneous correlation functions
 (\ref{eq.5.7.17})
do {\it not } depend on $\rho_0$. This is due to the fact that the second term of (\ref{eq.5.7.11}) does {\it not} depend on $f(q)$.

\vspace{0.3cm}

1.ii) Another interesting asymptotic regime to investigate is the one 
first studied by Torney and McConnell \cite{Torney}, where one considers initially very 
diluted systems, i.e. $\rho_0\approx \mu \ll 1$, but keeps the products $\epsilon\rho_0^2 t,\;
 \epsilon\rho_0^2 t_0$,
$\epsilon'\rho_0^2 t,\; \epsilon'\rho_0^2 t_0 $ fixed and finite, with 
 $\epsilon t, \epsilon t_0 \gg 1$ and $ \epsilon 't, \epsilon 't_0 \gg 1$ [and ${\widetilde c}>0$].

In this regime, the noninstantaneous two-point correlation functions ${\cal C}_{r}(t,t_0)$ are still 
dominated by the second term of (\ref{eq.5.7.11}) and thus the asymptotic ($v\neq 0$)
decay of ${\cal C}_{r}(t,t_0)$ is still given by (\ref{eq.5.7.17}).

The situation is however different for the {\it instantaneous} correlation functions [because the
 third and fourth term of (\ref{eq.5.7.11}) depend on $f(q)$].
With help of  (\ref{eq.5.7.18}), we
obtain ($r<\infty$): 
\begin{eqnarray}
\label{eq.5.7.21}
{\cal C}_{r}(t)\approx 
\rho_0^3 {\rm e}^{-4\epsilon't}
{\cal G}(\rho_{0}, r, \epsilon, \epsilon') \left[
\frac{1}{2(\rho_0^2 \pi {\widetilde c}t)^{1/2}}
-{\rm e}^{4\rho_{0}^{2}{\widetilde c}t}
{\rm Erfc}(2\rho_0 \sqrt{{\widetilde c}t})
\right],
\end{eqnarray}
where ${\rm Erfc}(z)$ denotes the usual complementary error function. The
 amplitude  $ {\cal G}(\rho_{0}, r, \epsilon, \epsilon') $
  has the following  form:
\begin{eqnarray}
\label{eq.5.7.22}
{\cal G}(\rho_{0}, r, \epsilon, \epsilon') &\equiv& (
A_0 -C_0)\left[\rho_{0}^{-2} -4r^2 +\epsilon/\epsilon'\right]
- B_{0}\sqrt{\epsilon'/\epsilon}\left[r/\rho_{0}^2 -10r(2r^2 +1)
 -\sqrt{\epsilon/\epsilon'}\right]
\end{eqnarray}

We now pass to the case of the DPA-model, where $\epsilon'=0$ and $b={\widetilde c}=\epsilon>0$.
Again, we distinguish two different regimes.

\vspace{0.3cm}

2.i) We first consider the regime where $bt\gg 1$ and $bt_0 \gg 1$.
In this situation, the main contribution to the long-time dynamics arises from
the fourth and the last terms of (\ref{eq.5.7.11}) and we obtain : 
\begin{eqnarray}
\label{eq.5.7.22}
{\cal C}_{r}(t, t_0;\epsilon'=0)\approx  
\frac{{\rm exp}\left[-\frac{(r-vt)^2}{bt} \frac{t_0+t}{2t_0 +t}\right]}{2\pi b\sqrt{t(t+2t_0)}} \left(1-\sqrt{\frac{t}{t+2t_0}} {\rm e}^{\frac{(r-vt)^2}{bt} \frac{t_0}{2t_0 +t}}\right)
\end{eqnarray}
 It has to be noticed that,
according to (\ref{eq.5.7.22}), when $r>0$ and $v>0$ (or, $r<0$ and $v<0$), 
${\cal C}_{r}(t, t_0)$ has a local maximum (a ``peak'') 
 at time $t_{p}\equiv r/v $. When $bt\gg bt_0 \gg 1$, we recover the result \cite{Schutz2,Grynberg-auto}: ${\cal C}_{r}(t,t_0; \epsilon'=0)\approx \frac{t_0 {\rm e}^{-(r-vt)^{2}/bt}}{2\pi bt^{2}}$.

In this regime, the main contribution to the {\it instantaneous}
 correlation function arises from the last term of (\ref{eq.5.7.11}), and
 we obtain the following result:
\begin{eqnarray}
\label{eq.5.7.23}
{\cal C}_{r}(t; \epsilon'=0)\approx
-1/{4\pi b t},
\end{eqnarray}
where the minus sign manifests the fact that the long-time dynamics of the DPA-model
is dominated by  {\it anticorrelation}, due to the pair-annihilation of the particles.
The results (\ref{eq.5.7.22})
and (\ref{eq.5.7.23}) and the fact that the latter  do {\it not}
depend on $\rho_0$ confirm, for random initial case, the universal character of
 the DPA-model in this regime.

\vspace{0.3cm}

2ii) We now consider the low-density regime of the DPA-model, where $\rho_0 \approx \mu \ll 1$
and $\epsilon t, \epsilon t_0\gg 1$, with $\epsilon \rho_0^2 t $ and 
$\epsilon \rho_0^2 t_0$ finite.
Also in this regime the main contribution to
 $ {\cal C}_{r}(t, t_0;\epsilon'=0)$ arises from  the fourth and the last terms
 of (\ref{eq.5.7.11})
and one  has the long-time behavior ($v=0$ and $r<\infty$):
\begin{eqnarray}
\label{eq.5.7.23.0}
{\cal C}_{r}(t,t_0; \epsilon'=0)\approx
\rho_{0}{\rm e}^{2\rho_0^2 b(t+2t_0)} {\rm Erfc}\left(
2\rho_{0}\sqrt{b(t_0+t/2)}\right)
\left\{\frac{1}{\sqrt{2\pi b t}} -
\rho_{0}{\rm e}^{2\rho_0^2 b(t+2t_0)} {\rm Erfc}\left(
2\rho_{0}\sqrt{b(t_0+t/2)}\right)
 \right\}
\end{eqnarray}

For the {\it instantaneous } correlation functions,  we obtain the following long-time behavior:
\begin{eqnarray}
\label{eq.5.7.23.1}
{\cal C}_{r}(t;\epsilon'=0)\approx 
-\rho_0^2 {\rm e}^{8\rho_0^2 bt}\left[{\rm Erfc}(2\rho_0 \sqrt{bt})\right]^2,
\end{eqnarray}
where the {\it anticorrelated} character of the DPA-model
 clearly appears.

\vspace{0.3cm}

Despite the fact that the  parameter
${\widetilde c}=\epsilon-\epsilon'$ can take negative values, so far we have always considered the
 case where ${\widetilde c}>0$. With help of the similarity transformation 
 ${\cal B}=\prod_{j=1}^{L}\sigma_{j}^{x}$, where 
$\sigma_{j}^x$ is the usual Pauli's matrix acting on the site $j$, we show that
 the case where ${\widetilde c}<0$ is directly related to the one where
 ${\widetilde c}>0$. In fact, according to  
 ${\cal B}$, the (free-fermion) DPAC stochastic Hamiltonian $H=H(b,{\widetilde c},v)$
is mapped onto $ {\cal B} H(b,{\widetilde c},v) {\cal B}^{-1}=
 H(b,-{\widetilde c}, -v)$ and the initial state $|\rho_{0}\rangle $  is mapped onto
$ {\cal B}|\rho_{0}\rangle = |1-\rho_{0}\rangle $.
Therefore, we have: ${\cal C}_{r}(t,t_{0})_{b,-{\widetilde c}, -v;\; \rho_{0}}=
{\cal C}_{r}(t,t_{0})_{ b,{\widetilde c},v;\; 1-\rho_{0}}.
$

\vspace{0.5cm}

In summary, in this work we  sketch a four-step procedure which allows 
the explicit computation of the multipoint and multitime correlation functions
of the free-fermion DPAC-model starting from  random (uncorrelated) initial states. We then 
specifically compute the noninstantaneous/instantaneous two-point correlation functions in the presence as well as in the absence of the pair-creation term. When all the reaction-rates are positive, the dynamics turns out to be non-universal and the long-time relaxation
is exponential (with a subdominant a power-law factor): the amplitude of the instantaneous two-point correlation functions depends on the initial density $\rho_0$ and is explictly determined.
In the absence of the pair creation, i.e. when $\epsilon'=0$ and $h+h'=\epsilon>0$, the
dynamics turns out to be universal (in the regime where $\rho_0$ 
is finite and $bt\gg 1$, $bt_0 \gg 1$) and there is a power-law relaxation.
The effect of the bias $v=h'-h\neq 0$, only appears in the noninstantaneous correlation functions and can be absorbed (for $\epsilon'\geq 0$) in a Galilean transformation, as noticed in \cite{Schutz2,Grynberg-auto}
in considering the DPA-model (in these previous works, $\epsilon'=0$). 

To close this work, it is natural to wonder what is the effect on the dynamics 
of the restriction $\gamma=0$. In fact, it is by now well established 
on the basis of numerous consistent numerical results
 \cite{Vilfan,Grynberg1-corr,Grynberg-auto,Barmo2}, 
and from  comparison with experiments \cite{Privman,Kuroda}, that
the results obtained for the free-fermion version of the DPAC-model give a {\it qualitative} 
 picture which is still valid  when  $\gamma \neq 0$. One can 
therefore expect that the results obtained in this work could have 
a general validity and, in particular, a direct relevance for recent interdisciplinary
 studies \cite{Vilfan}.  

\vspace{0.3cm}

We are grateful to P.-A. Bares, M. Michalakis
 and L. Zuppiroli for discussions. The support of the Swiss 
National Foundation is acknowledged.


\begin{thebibliography}{99}
\bibitem{Privman}
{\it Nonequilibrium Statistical Mechanics in One Dimension}, edited by V. Privman (Cambridge University Press, Cambridge, England, 1997).

\bibitem{Kuroda}
N. Kuroda, M. Nishida, Y.Tabata, Y. Wakabayashi and K. Sasaki, Phys. Rev. B {\bf 61}, 11217 (2000)

\bibitem{Vilfan}
A. Vilfan, E. Frey and F. Schwabl, Europhys. Lett. {\bf 56}, 420 (2001)

\bibitem{Schutzrev}
G.M. Sch\"utz, in {\it Phase Transitions and Critical Phenomena} edited by C. Domb and 
J. Lebowitz (Academic Press, London, 2000), Vol. 19



\bibitem{Lushnikov1}
A.A. Lushnikov, Sov. Phys. JETP {\bf 64}, 811 (1986)


\bibitem{Grynberg1-corr}
M.D. Grynberg, T.J. Newman and R.B. Stinchcombe, Phys. Rev. E {\bf 50 }, 957 (1994);M.D. Grynberg and R.B. Stinchcombe, Phys. Rev. Lett. {\bf 74 }, 1242 (1995);
M.D. Grynberg and R.B. Stinchcombe, Phys. Rev. E {\bf 52}, 6013 (1995);
M.D. Grynberg,  Phys. Rev. E {\bf 57}, 74 (1998)

%
%

\bibitem{Schutz2}
G. M. Sch\"utz, Phys. Rev. E {\bf 53}, 1475  (1996)

\bibitem{Schutz1}
G. M. Sch\"utz, J.Phys. A {\bf 28}, 3405  (1995)
%
\bibitem{Santos1}
J. Santos, G.M. Sch\"utz, and R. B. Stinchcombe, J. Chem. Phys. {\bf 105}, 2399 (1996)
%
\bibitem{Mobar2}
M. Mobilia and P.-A. Bares, Phys. Rev. E {\bf 63}, 056112 (2001)
%
\bibitem{Grynberg-auto}
M. D. Grynberg and R. B. Stinchcombe, Phys. Rev. Lett. {\bf 76}, 851 (1996)
%
\bibitem{Barmo1}
P.-A. Bares and M. Mobilia, Phys. Rev. E {\bf 59}, 1996 (1999)
%
\bibitem{Oliveira}
M. J. de Oliveira, Phys. Rev. E {\bf 60}, 2563 (1999)
%
%
\bibitem{Barmo2}
P.-A. Bares and M. Mobilia, Phys. Rev. Lett. {\bf 83}, 5214 (1999);
 P.-A. Bares and M. Mobilia, Phys. Rev. Lett. {\bf 85}, 893 (2000);
S.C. Park, J.-M. Park and D. Kim, Phys. Rev. Lett. {\bf 85}, 892 (2000);
S.C. Park, J.-M. Park and D. Kim, Phys. Rev. E {\bf 63}, 057102 (2001)


\bibitem{Torney}
D. C. Torney and M. McConnell, J. Phys. Chem. {\bf 87}, 1941 (1983)

\bibitem{Mobar3}
M. Mobilia, ``Non-equilibrium Systems in Statistical Mechanics: Some Exactly Solvable Reaction-diffusion Models'', Thesis EPFL n.2552 (April 2002).

\bibitem{Derrida1}
B. Derrida and R. Zeitak, Phys. Rev. E {\bf 54}, 2513 (1996)


\bibitem{Derrida2}
B. Derrida, V. Hakim, and V. Pasquier, J. Stat. Phys. {\bf 85}, 763 (1996)

\bibitem{Derrida3}
B. Derrida, V. Hakim, and V. Pasquier, Phys. Rev. Lett. {\bf 75}, 751 (1995)

\end{thebibliography}
\end{document}